\documentclass[a4paper,10pt]{article}

\usepackage{amsfonts}
\usepackage{amsmath}
\usepackage[all]{xy}
\usepackage{verbatim}
\usepackage{rotating}

\usepackage[ps2pdf]{hyperref}
\hypersetup{pdftitle = {Complete analysis of extensions of $D(n)_1$ permutation orbifolds}}
\hypersetup{pdfauthor = {Michele Maio, Bert Schellekens}}
\hypersetup{pdfsubject = {CFT-String Theory}}
\hypersetup{pdfkeywords = {Orbifolds, Extensions, Fixed Point Resolution}}

\numberwithin{equation}{section}
\begin{document}

\title{Complete analysis of extensions of $D(n)_1$ permutation orbifolds}
\author{M. Maio$^{1}$ and A.N. Schellekens$^{1,2,3}$\\~\\~\\
\\
$^1$Nikhef Theory Group, Amsterdam, The Netherlands
\\
~\\
$^2$IMAPP, Radboud Universiteit Nijmegen, The Netherlands
\\
~\\
$^3$Instituto de F\'\i sica Fundamental, CSIC, Madrid, Spain}

\maketitle

\begin{abstract}
We give the full set of $S$ matrices for extensions of $D(n)_1$ permutation orbifolds,
extending our previous work
to the yet unknown case of integer spin spinor currents. The main tool is triality of $SO(8)$. We also provide 
fixed point resolution matrices for spinor currents of $D(n)_1$ permutation orbifolds with $n$ even and not multiple of four, where the spinor currents have half-integer spin.
\end{abstract}

\clearpage
\tableofcontents

\section{Introduction}
In our previous paper \cite{Maio:2009kb} we studied the structure of order-two simple currents in permutation orbifolds in two-dimensional conformal field theories \cite{Belavin:1984vu}. The main tool was the BHS $S$ matrix for the permutation orbifold \cite{Klemm:1990df,Borisov:1997nc}. In general they can only be generated from diagonal fields that correspond to simple currents in the mother theory, while their fixed points can come from both the untwisted (diagonal and off-diagonal) and twisted sector. In the same paper we also considered extensions of the permutation orbifold and fixed point resolution. 

Simple current extensions are useful tools in conformal field theories and string theory 
\cite{Schellekens:1989am,Schellekens:1989dq,Intriligator:1989zw} but their
modular transformation matrices are often
quite non-trivial due to fixed points \cite{Schellekens:1999yg,Schellekens:1989uf}. In \cite{Maio:2009kb} we derived $S$ matrices for  extensions in the case of $SU(2)_2$, $B(n)_1$ and $D(n)_1$ WZW models \cite{Knizhnik:1984nr,Gepner:1986wi}. This was completely done for the first two models but only partially for the $D(n)_1$. In fact, we provided the $S$ matrix for the omnipresent integer spin simple currents for any value of $n$, but sometimes additional currents appear in the $D(n)_1$ model whose fixed points must be resolved as well, in order to use them as extensions. Generically fixed points can arise for integer spin and half-integer spin simple currents \cite{Schellekens:1990xy}. We will see that this happens for particular ranks of $D(n)_1$ where they must be resolved. In this paper we address those additional problems, providing a complete picture for the fixed point resolution in $D(n)_1$ permutation orbifold.

Explicitly, there are two interesting situations where fixed points can occur and that we have not studied so far. When $n$ is multiple of four, $n=4p$ with $p\in \mathbb{Z}$, there are additional integer-spin simple currents coming from the two spinor representations of the $D(n)_1$ WZW model. The spinor fields have weight $h=\frac{n}{8}$ and their symmetric and anti-symmetric representations in the $D(n)_1$ permutation orbifold have weight $h=\frac{n}{4}$. Similarly, when $n=4p+2$, the same two spinor currents generate half-integer spin simple currents in the $D(n)_1$ permutation orbifold. 
Although the latter cannot be used to extend the chiral algebra, they can be used in combination with half-integer spin currents of
another factor in a tensor product. For example, one may tensor the permutation orbifold with an Ising model, and consider 
the product of the half-integer spin current of the $D(n)_1$ permutation orbifold and the Ising spin field. This is not just of academic interest.
Extended tensor products of rational conformal field theories are an important tool in explicit four-dimensional string constructions, and
in the  vast majority of cases one encounters fixed points. For this reason the fixed point resolution matrices we determine
here and in our previous paper $\cite{Maio:2009kb}$ have a range of applicability far beyond the special cases used here to 
determine them. 

From previous works \cite{Fuchs:1996dd}, we know that resolving the fixed points is the same as finding a set of $S^J$ matrices, one matrix for each current $J$, acting on the fixed points. The $S^J$ matrices must be unitary and must satisfy the modular constraint $(S^J)^2=(S^JT^J)^3$, where $T^J$ is the $T$ matrix of the extended theory restricted to the fixed points; moreover, for order-two currents, the $S^J$'s must be symmetric. The $S$ matrix of the extended theory is then computed as a Fourier-like transform of the $S^J$ matrices \cite{Fuchs:1996dd}: it has to be unitary, modular invariant and should give rise to non-negative integer fusion coefficients obtained by the Verlinde formula \cite{Verlinde:1988sn}. These are non-trivial tests for a good $S$ matrix.

There is no known algorithm for determining these matrices in generic rational CFT's, even if their matrix $S$ is known. In WZW-based
models (WZW extensions and coset CFT's) one can make use of foldings of Dynking diagrams 
\cite{Fuchs:1995zr} to compute the matrices $S^J$. In $\cite{Maio:2009kb}$ we made use of the fact that the extension currents
had spin 1 and led to
identifiable CFT's. This method will not work here except in the special case of $D(4)$, where the spinor currents have spin 1.
In that case one can make use of triality of $SO(8)$ to determine the missing fixed point resolution matrices. Although triality does not extend
to larger ranks, it turns out that in the other cases the fixed point spectrum is sufficiently similar to allow us to make a general {\it ansatz}.

The plan of the paper is as follows.\\
In section \ref{D4p permutation orbifolds} we describe the $D(4p)_1$ permutation orbifolds extended by the two spinor currents and resolve the fixed points. In the special case $p=1$ we 
use triality of $SO(8)$ to determine the  set of $S^J$ matrices. From the case $p=1$ is indeed possible to generalize the result to arbitrary values of $p$.\\
In section \ref{D4p+2 permutation orbifolds} we repeat the procedure for $D(4p+2)_1$ permutation orbifolds. We can be fast here since a very few changes are sufficient to write down consistent $S^J$ matrices.\\
We conclude by illustrating open questions and future directions.

\section{$D(4p)_1$ permutation orbifolds}
\label{D4p permutation orbifolds}
By permutation orbifold we mean the procedure of taking the tensor product of a given conformal field theory (that sometimes we will call the mother theory) with itself (we restrict ourselves to two factors in the tensor product, even if it is possible to generalize the product to more than two factors \cite{Bantay:1997ek,Bantay:1999us}) and modding out the resulting theory with respect to the permutation symmetry which exchanges the two factors. All the details of the orbifold theory are known from the work of BHS \cite{Borisov:1997nc}. Once the permutation orbifold is given, we may extend it by any of its integer spin simple currents to derive new theories. The presence of simple current fixed points makes life difficult, because the new extended $S$ matrix is not trivially known in terms of the one of the mother theory, but requires the knowledge of a set  of $S^J$ matrices, one for each simple current. Here we start with the $D(n)_1$ WZW model as mother theory and focus on the spinor currents that for even rank $n$ can have (half-)integer spin. This will complete the analysis initiated in \cite{Maio:2009kb}.

Let us fix our notation. The $D(n)_1=SO(N)_1$, $N=2n$, series has central charge $c=\frac{N}{2}$ and four primary fields $\phi_i$ with weight $h_i=0,\frac{N}{16},\frac{1}{2},\frac{N}{16}$ ($i=0,1,2,3$ respectively).
The $S$ matrix is given in table \ref{table S_Dn_1}.
\begin{table}[ht]
\caption{$S$ matrix for $D(n)_1$}
\centering
\begin{tabular}{c|c c c c}
\hline \hline\\
$S_{D(n)_1}$ & $h=0$ &  $h=\frac{N}{16}$ & $h=\frac{1}{2}$ & $h=\frac{N}{16}$ \\ 
\hline &&&\\
$h=0$             & $\frac{1}{2}$        & $\frac{1}{2}$         & $\frac{1}{2}$ &     $\frac{1}{2}$ \\
$h=\frac{N}{16}$  & $\frac{1}{2}$        & $\frac{(-i)^n}{2}$    & $-\frac{1}{2}$ &    $-\frac{(-i)^n}{2}$ \\
$h=\frac{1}{2}$   & $\frac{1}{2}$        & $-\frac{1}{2}$        & $\frac{1}{2}$ &     $-\frac{1}{2}$ \\
$h=\frac{N}{16}$  & $\frac{1}{2}$        & $-\frac{(-i)^n}{2}$   & $-\frac{1}{2}$ &    $\frac{(-i)^n}{2}$
\end{tabular}
\label{table S_Dn_1}
\end{table}

All the four fields of the $D(n)_1$ series are simple currents. 
In the permutation orbifold, they give rise to four integer spin simple currents, namely $(0,0)$, $(0,1)$, $(2,0)$ and $(2,1)$, and to four non-necessarily-integer spin simple currents, namely $(1,0)$, $(1,1)$, $(3,0)$ and $(3,1)$. For $n$ multiple of four, these latter currents have also integer spin. In \cite{Maio:2009kb} we focused on the former set. Here we want to study the latter, coming from the spinor representations $i=1,3$ of the $D(n)_1$ model.

There are already a few observations that we can make. First of all, there exists an automorphism that exchanges the fields $\phi_1$ and $\phi_3$. This will have the consequence that the permutation theories extended by the currents $(1,0)$ and $(3,0)$ will be isomorphic\footnote{The fields $\phi_1$ and $\phi_3$ also have same $P$-matrix entries. In fact, the $P$ matrix for $n=4p$ is
\begin{equation}
P=\left(
\begin{array}{cccc}
(-1)^p &  0  &    0       & 0 \\
0      &  1  &    0       & 0 \\
0      &  0  & (-1)^{p+1} & 0 \\
0      &  0  &    0       & 1
\end{array}
\right)\,. \nonumber
\end{equation}
We recall that the $P$ matrix, $P=\sqrt{T}ST^2S\sqrt{T}$, first introduced in \cite{Pradisi:1995qy}, enters the BHS formulas \cite{Borisov:1997nc} for the $S$ matrix of the permutation orbifold in the twisted sector.} (the fields having same weights and the two theories having equal central charge); this holds as well as for the extensions by $(1,1)$ and $(3,1)$. Secondly, when $n$ is multiple of four, i.e. $n=4p$ with $p\in\mathbb{Z}$, the $S$ matrix of the mother $D(n)_1$ theory is the same for every $p$. This will have the consequence that the fusion rules of these current in the permutation orbifolds are the same for every value of $p$. Putting these two observations together, we conclude that for $n=4p$ there will be only two universal $S^J$ matrices to determine\footnote{They will in general depend on $p$ through a phase in order to satisfy modular invariance, since the $T$ matrix depends on $p$.}.

Let us illustrate these points with the explicit construction. Consider\footnote{The case $n=4$, that we will consider extensively later, is very interesting since it corresponds to $SO(8)_1$ where, due to triality, three out of four fields have equal weight. The extensions by the currents $(1,\psi)$, $(2,\psi)$ and $(3,\psi)$ must produce the same result. The extension by $(2,\psi)$ is already known from \cite{Maio:2009kb} and from what we said before we also know that the extensions by $(1,\psi)$ and $(3,\psi)$ are equal. Indeed one can check that the extension of the permutations orbifold by $(1,0)$ is
\begin{equation}
D(4)_1\times D(4)_1/\mathbb{Z}_2 = (SU(8)_1\times U(1)_{128})_{(4,16\cdot x)}
\end{equation}
for odd integer $x$. Also, the extension by $(1,1)$ gives a theory which is almost isomorphic to (in the sense of having same weights and same central charge but not dimensions of) the tensor product of two $D(4)_1$'s.} the case with arbitrary $n=4 p$. The $D(n)_1$ weights are then $h=0,\frac{n}{8},\frac{1}{2},\frac{n}{8}$ and the orbit structure under the additional integer-spin simple currents (all with $h=\frac{n}{4}=p$) is as follows.

\begin{tabular}{l l l l}
&&&\\
$J\equiv(1,0)$ & \underline{Fixed points} &\phantom{$J\equiv(1,0)$}& \underline{Length-2 orbits}\\
& $(\phi_0,\phi_1)$, \,$h=\frac{n}{8}$              && $\Big( (0,0),(1,0) \Big)$,\, $h=0$\\
& $(\phi_2,\phi_3)$, \,$h=\frac{n}{8}+\frac{1}{2}$  && $\Big( (0,1),(1,1) \Big)$,\, $h=1$\\
& $\widehat{(0,0)}$,\, $h=\frac{n}{16}$             && $\Big( (2,0),(3,0) \Big)$,\, $h=1$\\
& $\widehat{(0,1)}$,\, $h=\frac{n}{16}+\frac{1}{2}$ && $\Big( (2,1),(3,1) \Big)$,\, $h=1$\\
& $\widehat{(1,0)}$,\, $h=\frac{n}{8}$              &&\\
& $\widehat{(1,1)}$,\, $h=\frac{n}{8}+\frac{1}{2}$  &&\\
&&&\\
\end{tabular}

\begin{tabular}{l l l l}
$J\equiv(1,1)$ & \underline{Fixed points} &\phantom{$J\equiv(1,1)$}& \underline{Length-2 orbits}\\
& $(\phi_0,\phi_1)$, \,$h=\frac{n}{8}$                          && $\Big( (0,0),(1,1) \Big)$,\, $h=0$\\
& $(\phi_2,\phi_3)$, \,$h=\frac{n}{8}+\frac{1}{2}$              && $\Big( (0,1),(1,0) \Big)$,\, $h=1$\\
& $\widehat{(2,0)}$,\, $h=\frac{n}{16}+\frac{1}{4}$             && $\Big( (2,0),(3,1) \Big)$,\, $h=1$\\
& $\widehat{(2,1)}$,\, $h=\frac{n}{16}+\frac{1}{4}+\frac{1}{2}$ && $\Big( (2,1),(3,0) \Big)$,\, $h=1$\\
& $\widehat{(3,0)}$,\, $h=\frac{n}{8}$                          &&\\
& $\widehat{(3,1)}$,\, $h=\frac{n}{8}+\frac{1}{2}$              &&\\
&&&\\
\end{tabular}

\begin{tabular}{l l l l}
$J\equiv(3,0)$ & \underline{Fixed points} &\phantom{$J\equiv(3,0)$}& \underline{Length-2 orbits}\\
& $(\phi_0,\phi_3)$, \,$h=\frac{n}{8}$                && $\Big( (0,0),(3,0) \Big)$,\, $h=0$\\
& $(\phi_1,\phi_2)$, \,$h=\frac{n}{8}+\frac{1}{2}$    && $\Big( (0,1),(3,1) \Big)$,\, $h=1$\\
& $\widehat{(0,0)}$,\, $h=\frac{n}{16}$               && $\Big( (1,0),(2,0) \Big)$,\, $h=1$\\
& $\widehat{(0,1)}$,\, $h=\frac{n}{16}+\frac{1}{2}$    && $\Big( (1,1),(2,1) \Big)$,\, $h=1$\\
& $\widehat{(3,0)}$,\, $h=\frac{n}{8}$                &&\\
& $\widehat{(3,1)}$,\, $h=\frac{n}{8}+\frac{1}{2}$    &&\\
&&&\\
\end{tabular}

\begin{tabular}{l l l l}
$J\equiv(3,1)$ & \underline{Fixed points} &\phantom{$J\equiv(3,1)$}& \underline{Length-2 orbits}\\
& $(\phi_0,\phi_3)$, \,$h=\frac{n}{8}$                           && $\Big( (0,0),(3,1) \Big)$,\, $h=0$\\
& $(\phi_1,\phi_2)$, \,$h=\frac{n}{8}+\frac{1}{2}$               && $\Big( (0,1),(3,0) \Big)$,\, $h=1$\\
& $\widehat{(1,0)}$,\, $h=\frac{n}{8}$                           && $\Big( (1,0),(2,1) \Big)$,\, $h=1$\\
& $\widehat{(1,1)}$,\, $h=\frac{n}{8}+\frac{1}{2}$               && $\Big( (1,1),(2,0) \Big)$,\, $h=1$\\
& $\widehat{(2,0)}$,\, $h=\frac{n}{16}+\frac{1}{4}$              &&\\
& $\widehat{(2,1)}$,\, $h=\frac{n}{16}+\frac{1}{4}+\frac{1}{2}$  &&\\
&&&\\
\end{tabular}

\noindent Note that in going from the fixed points of $(1,\psi)$ to $(3,\psi)$, the fields $\phi_1$ and $\phi_3$ get interchanged: this provides isomorphic sets of fields in the extensions.

The fixed points get splitted into two fields in the extended permutation orbifold and hence all the theories above admit $2\cdot 6+4=16$ fields. By changing $n=4p$, the weights of the orbits and the ones of the fixed points might change, but as we said there are a few things that remain invariant, namely: 1) the fact that the extension by the current $(1,0)$ (resp. $(1,1)$) is isomorphic (up to field reordering) to the one by $(3,0)$ (resp. $(3,1)$), as it can be seen by looking at the weights of the extended fields; 2) the orbit and fixed-point structure (i.e. the fusion rules of the currents with any other field in the permutation orbifold) remains the same for arbitrary $p$; this has the consequence that we will have to determine only two $S^J$ matrices instead of four.

\subsection{$S^J$ matrices for $D(4p)_1$ permutation orbifolds}
We have already noticed that there are in practice only two $S^J$ matrices to determine for the four above-mentioned integer-spin simple currents. So here we derive $S^{J=(1,0)}$ and $S^{J=(1,1)}$; $S^{J=(3,0)}$ and $S^{J=(3,1)}$ are equal to the former two, after proper field ordering.

It is instructive to start with the $D(4)_1$ $(p=1)$ case. $SO(8)_1$ is special in the sense that the three non-trivial representations, i.e. the vector ${\bf 8_v}$ and the two spinors ${\bf 8_s}$ and ${\bf 8_c}$, have same weight ($h=\frac{1}{2}$) and same multiplicity (dim$=8$) and can be mapped into each other. This property of $SO(8)$ is triality.

Let us now work out the $S^J$ matrices corresponding to the two integer-spin simple currents $J=(1,0)$ and $J=(1,1)$. The extension by $(1,0)$ of the permutation orbifold is isomorphic to an extension of the tensor product of an $SU(8)$ and a $U(1)$ factor as in \cite{Maio:2009kb}:
\begin{equation}
(D(4)_1 \times D(4)_1/ \mathbb{Z}_2)_{(1,0)}=(SU(8)_1 \times U(1)_{128})_{(4,16)}\,,
\end{equation}
while the extension by $(1,1)$ is isomorphic to the tensor product $D(4)_1 \times D(4)_1$. This is exactly what happened for the already known currents $(2,\psi)$ \cite{Maio:2009kb}; in fact, due to triality of $SO(8)$, the three theories extended by $(1,\psi)$ $(2,\psi)$ $(3,\psi)$ must be the same.

\subsubsection{$J=(1,0)$}
We use the main formula of \cite{Fuchs:1996dd}
\begin{equation}
\label{MainFormula}
\tilde{S}_{(a,i)(b,j)}=\frac{|G|}{\sqrt{|U_a||S_a||U_b||S_b|}}\sum_{J\in G}\Psi_i(J) S^J_{ab} \Psi_j(J)^{\star}
\end{equation}
as done in \cite{Maio:2009kb} to derive the $S^J$ matrix from the knowledge of the extended matrix $\tilde{S}$ and the permutation orbifold matrix $S^{(0,0)}\equiv S^{BHS}$. The prefactor in (\ref{MainFormula}) is a group theoretical factor and the $\Psi_i$'s are the group characters. Our field convention to distinguish between the two splitted fixed points is:
\begin{eqnarray}
(\phi_0,\phi_1) \,\, \longrightarrow&  (1,4) & \&\qquad (7,124) \nonumber \\
(\phi_2,\phi_3) \,\, \longrightarrow&  (1,116) & \&\qquad (3,124) \nonumber \\
\widehat{(0,0)} \,\, \longrightarrow&  (0,120) & \&\qquad (0,8) \nonumber \\
\widehat{(0,1)} \,\, \longrightarrow&  (6,0) & \&\qquad (2,0) \nonumber \\
\widehat{(1,0)} \,\, \longrightarrow&  (7,4) & \&\qquad (1,124) \nonumber \\
\widehat{(1,1)} \,\, \longrightarrow&  (1,12) & \&\qquad (3,4) \nonumber \\
&&\nonumber
\end{eqnarray}
where $(s,u)$ denote a field in the extended theory ($s\equiv s+8$, $u\equiv u+128$). Observe that field one and field two correspond to complementary orbits as explained in \cite{Maio:2009kb}. We obtain the matrix in table \ref{table S^J=10_p=1} for the $(D4_1\times D4_1/\mathbb{Z}_2)_{(1,0)}$ orbifold. We denote it by $S^J_{D4}$ for reasons that will become clear later.
\begin{table}[ht]
\caption{Fixed point Resolution: Matrix $S^{J\equiv (1,0)}_{D4}$}
\centering
\begin{tabular}{c|c c c c c c c}
\hline \hline\\
$S^{J\equiv (1,0)}_{D4}$ & $(\phi_0,\phi_1)$ & $(\phi_2,\phi_3)$ & $\widehat{(0,0)}$ & $\widehat{(0,1)}$ & $\widehat{(1,0)}$ & $\widehat{(1,1)}$ \\ 
\hline &&&\\
$(\phi_0,\phi_1)$   & $0$  & $0$     & $\frac{i}{2}$  & $-\frac{i}{2}$ & $-\frac{i}{2}$ & $-\frac{i}{2}$ \\
$(\phi_2,\phi_3)$   & $0$  & $0$     & $\frac{i}{2}$  & $-\frac{i}{2}$ & $\frac{i}{2}$  & $\frac{i}{2}$ \\
$\widehat{(0,0)}$   & $\frac{i}{2}$  & $\frac{i}{2}$  & $0$  & $0$ & $\frac{i}{2}$  & $-\frac{i}{2}$ \\
$\widehat{(0,1)}$   & $-\frac{i}{2}$ & $-\frac{i}{2}$ & $0$  & $0$ & $\frac{i}{2}$  & $-\frac{i}{2}$ \\
$\widehat{(1,0)}$   & $-\frac{i}{2}$ & $\frac{i}{2}$  & $\frac{i}{2}$  & $\frac{i}{2}$  & $0$  & $0$ \\
$\widehat{(1,1)}$   & $-\frac{i}{2}$ & $\frac{i}{2}$  & $-\frac{i}{2}$ & $-\frac{i}{2}$ & $0$  & $0$ \\
\end{tabular}
\label{table S^J=10_p=1}
\end{table}

One can check that this matrix is unitary ($S^J (S^J)^\dagger=1$) and modular invariant ($(S^J)^2=(S^J T^J)^3$, where $T^J$ is the $T$ matrix restricted to the fixed points) and gives non-negative integer fusion coefficients. Moreover, one can see that unitarity and modular invariance are preserved for $p=1$ mod $4$: then this matrix can be used also in these situations.

Observe that rescaling the $S^J$ matrix by a phase does not destroy unitarity but it does affect modular invariance. By a suitable choice of the phase, it is possible to make a modular invariant matrix out of $S^{(1,0)}_{D4}$ valid for all $p$. The correct choice is:
\begin{equation}
\label{S10p}
\boxed{S^{(1,0)}= (-i)^{p-1} \cdot S^{(1,0)}_{D4}= e^{-\frac{i\pi}{4}(m-2)} \cdot S^{(1,0)}_{D4}}
\end{equation}
which will use for any value of $p$. Here $m=2p$ is an even integer such that $D_{2m}\equiv D_{4p}$. This is again unitary, modular invariant and gives non-negative integer fusion coefficients.

Let us make a final comment. What happens when we shift $p\rightarrow p+1$? Under this shift, the fixed point weights change differently. In particular, for the current $(1,0)$ the shifts are $h\rightarrow h+\{\frac{1}{2},\frac{1}{2},\frac{1}{4},\frac{1}{4},\frac{1}{2},\frac{1}{2}\}$. The $T^{(1,0)}$ matrix then changes as $T^{(1,0)}\rightarrow e^{-\frac{2\pi i}{3}}\,{\rm diag}(-1,-1,i,i,-1,-1)\cdot T^{(1,0)}$ (the phase in front coming from the central charge), while the $S^{(1,0)}$ takes a phase, $S^{(1,0)}\rightarrow -iS^{(1,0)}$. These changes are such that modular invariance is still preserved for every $p$.

\subsubsection{$J=(1,1)$}
For this current, recall that
\begin{equation}
(D(4)_1 \times D(4)_1/ \mathbb{Z}_2)_{(1,1)}\sim D(4)_1 \times D(4)_1\,.
\end{equation}
The splitted fixed points correspond to fields in the tensor product theory. We choose conventionally the following scheme, but a few other choices are also possible.
\begin{eqnarray}
(\phi_0,\phi_1) \,\, \longrightarrow&  \phi_0 \otimes \phi_1  & \&\qquad \phi_1 \otimes \phi_0 \nonumber \\
(\phi_2,\phi_3) \,\, \longrightarrow&  \phi_2 \otimes \phi_3  & \&\qquad \phi_3 \otimes \phi_2 \nonumber \\
\widehat{(2,0)} \,\, \longrightarrow&  \phi_0 \otimes \phi_2  & \&\qquad \phi_2 \otimes \phi_0 \nonumber \\
\widehat{(2,1)} \,\, \longrightarrow&  \phi_1 \otimes \phi_3  & \&\qquad \phi_3 \otimes \phi_1 \nonumber \\
\widehat{(3,0)} \,\, \longrightarrow&  \phi_0 \otimes \phi_3  & \&\qquad \phi_3 \otimes \phi_0 \nonumber \\
\widehat{(3,1)} \,\, \longrightarrow&  \phi_1 \otimes \phi_2  & \&\qquad \phi_2 \otimes \phi_1 \nonumber \\
&&\nonumber
\end{eqnarray}
Now our strategy to compute $S^J_{D4}$ is as follow. We first go to the isomorphic tensor product theory and use
\begin{equation}
S^J_{(mn)(pq)}=S_{mp}S_{nq}-S_{mq}S_{np}
\end{equation}
as derived in \cite{Maio:2009kb} to compute the $S^J$ matrix there and then we go back to the extended permutation orbifold using the field map. We obtain the $S^J$ matrix as in table \ref{table S^J=11_p=1}.
\begin{table}[ht]
\caption{Fixed point Resolution: Matrix $S^{J\equiv (1,1)}_{D4}$}
\centering
\begin{tabular}{c|c c c c c c c}
\hline \hline\\
$S^{J\equiv (1,1)}_{D4}$ & $(\phi_0,\phi_1)$ & $(\phi_2,\phi_3)$ & $\widehat{(2,0)}$ & $\widehat{(2,1)}$ & $\widehat{(3,0)}$ & $\widehat{(3,1)}$ \\ 
\hline &&&\\
$(\phi_0,\phi_1)$   & $0$  & $0$     & $-\frac{1}{2}$ & $-\frac{1}{2}$ & $-\frac{1}{2}$ & $-\frac{1}{2}$ \\
$(\phi_2,\phi_3)$   & $0$  & $0$     & $-\frac{1}{2}$ & $-\frac{1}{2}$ & $\frac{1}{2}$  & $\frac{1}{2}$ \\
$\widehat{(2,0)}$   & $-\frac{1}{2}$ & $-\frac{1}{2}$ & $0$  & $0$ & $-\frac{1}{2}$ & $\frac{1}{2}$ \\
$\widehat{(2,1)}$   & $-\frac{1}{2}$ & $-\frac{1}{2}$ & $0$  & $0$ & $\frac{1}{2}$  & $-\frac{1}{2}$ \\
$\widehat{(3,0)}$   & $-\frac{1}{2}$ & $\frac{1}{2}$  & $-\frac{1}{2}$ & $\frac{1}{2}$  & $0$  & $0$ \\
$\widehat{(3,1)}$   & $-\frac{1}{2}$ & $\frac{1}{2}$  & $\frac{1}{2}$  & $-\frac{1}{2}$ & $0$  & $0$ \\
\end{tabular}
\label{table S^J=11_p=1}
\end{table}

The $S^J$ matrix obtained in this way for $(D(4)_1 \times D(4)_1/ \mathbb{Z}_2)_{(1,1)}$ is unitary and modular invariant, so it is a good matrix for the extended theory. Moreover, this $S^J$ matrix is a good (i.e. unitary and modular invariant) matrix also for $p=1$ mod $4$.

In order to make this matrix modular invariant for any $p$, we again multiply by a phase. The choice is the same as before:
\begin{equation}
\label{S11p}
\boxed{S^{(1,1)}= (-i)^{p-1} \cdot S^{(1,1)}_{D4}= e^{-\frac{i\pi}{4}(m-2)} \cdot S^{(1,1)}_{D4}}
\end{equation}
which will use for any value of $p$. This is again unitary, modular invariant and gives non-negative integer fusion coefficients. Again, the shift $n\rightarrow n+16$ changes $S^{(1,1)}$ by a phase and $T^{(1,1)}$ in a more complicated way, but both always in a modular invariant fashion.

One can check formulas (\ref{S10p}) and (\ref{S11p}) in many explicit examples. For instance, one can see that they have good properties by looking at a few values of $p$, but also considering tensor products like $D(8)_1\times D(12)_1$ or $D(8)_1\times D(16)_1$ and extending with many current combinations $(J_1,J_2)$, where $J_1$ belongs to the first factor and $J_2$ to the second factor. In every example, the fusion rules give non-negative integer coefficients.

\section{$D(4p+2)_1$ permutation orbifolds}
\label{D4p+2 permutation orbifolds}
So far we have not addressed half-integer spin simple currents. They might also admit fixed points that must be resolved in the extended theory. This happens for the $D(n)_1$ permutation orbifolds with $n=4p+2$. In fact, the four currents $(1,\psi)$ and $(3,\psi)$, with $\psi =0,1$, will have weight $h=\frac{2p+1}{2}$ and will admit fixed points. The orbit structure is in this case with $n=4p+2$ very similar to the previous situation with $n=4p$, except for the fact that the twisted fields get reshuffled. The fixed point structure is as following. Observe that this is very similar to the structure for the previous case $n=4p$.

\begin{tabular}{l l l l}
&&&\\
$J\equiv(1,0)$ & \underline{Fixed points} &$J\equiv(3,0)$& \underline{Fixed points}\\
& $(\phi_0,\phi_1)$,\,$h=\frac{n}{8}$               && $(\phi_0,\phi_3)$, \,$h=\frac{n}{8}$\\
& $(\phi_2,\phi_3)$,\,$h=\frac{n}{8}+\frac{1}{2}$   && $(\phi_1,\phi_2)$, \,$h=\frac{n}{8}+\frac{1}{2}$\\
& $\widehat{(2,0)}$,\,$h=\frac{n}{16}+\frac{1}{4}$            && $\widehat{(2,0)}$,\, $h=\frac{n}{16}+\frac{1}{4}$\\
& $\widehat{(2,1)}$,\,$h=\frac{n}{16}+\frac{1}{4}+\frac{1}{2}$&& $\widehat{(2,1)}$,\,$h=\frac{n}{16}+\frac{1}{4}+\frac{1}{2}$\\
& $\widehat{(1,0)}$,\, $h=\frac{n}{8}$             && $\widehat{(3,0)}$,\, $h=\frac{n}{8}$\\
& $\widehat{(1,1)}$,\, $h=\frac{n}{8}+\frac{1}{2}$ && $\widehat{(3,1)}$,\, $h=\frac{n}{8}+\frac{1}{2}$\\
&&&\\
\end{tabular}

\begin{equation}
\label{D 4p+2 scheme}
\end{equation}

\begin{tabular}{l l l l}
&&&\\
$J\equiv(1,1)$ & \underline{Fixed points} &$J\equiv(3,1)$& \underline{Fixed points}\\
& $(\phi_0,\phi_1)$, \,$h=\frac{n}{8}$              && $(\phi_0,\phi_3)$, \,$h=\frac{n}{8}$\\
& $(\phi_2,\phi_3)$, \,$h=\frac{n}{8}+\frac{1}{2}$  && $(\phi_1,\phi_2)$, \,$h=\frac{n}{8}+\frac{1}{2}$\\
& $\widehat{(0,0)}$,\, $h=\frac{n}{16}$             && $\widehat{(0,0)}$,\, $h=\frac{n}{16}$\\
& $\widehat{(0,1)}$,\, $h=\frac{n}{16}+\frac{1}{2}$ && $\widehat{(0,1)}$,\, $h=\frac{n}{16}+\frac{1}{2}$\\
& $\widehat{(3,0)}$,\, $h=\frac{n}{8}$              && $\widehat{(1,0)}$,\, $h=\frac{n}{8}$\\
& $\widehat{(3,1)}$,\, $h=\frac{n}{8}+\frac{1}{2}$  && $\widehat{(1,1)}$,\, $h=\frac{n}{8}+\frac{1}{2}$\\
&&&\\
\end{tabular}

Again, the current $(1,0)$ (resp. $(1,1)$) generates the same fixed points as the current $(3,0)$ (resp. $(3,1)$), hence we have to determine only two, instead of four, $S^J$ matrices, since $S^{(1,\psi)}=S^{(3,\psi)}$, with $\psi=0,1$. Actually the study of the previous section helps us a lot, since it is easy to generate unitary and modular invariant matrices out of two matrices \emph{numerically} equal to the two $S^J_{D4}$ matrices of tables \ref{table S^J=10_p=1} and \ref{table S^J=11_p=1} with the fields ordered as above. More tricky is to check that also the fusion coefficients are non-negative integers if these currents are used in chiral algebra extensions.

The more sensible choice is the following. Let us have a closer look at the fixed point structure of the $n=4p$ and the $n=4p+2$ cases. They are very similar, but not quite. The weights of the fixed points of the current $(1,0)$ in the $n=4p$ case have the same expression as the weights of the fixed points of the current $(1,1)$ in the $n=4p+2$ case, and similarly for the $(3,\psi)$ current. So a natural guess for the $S^J$ matrices would involve interchanging the matrices in tables \ref{table S^J=10_p=1} and \ref{table S^J=11_p=1}. Equivalently, symmetric and anti-symmetric representations are interchanged in going from $n=4p$ to $n=4p+2$. Hence, we would expect $S^{(1,0)} \sim S^{(1,1)}_{D4}$ and $S^{(1,1)} \sim S^{(1,0)}_{D4}$. This is indeed the case. The unitary and modular invariant\footnote{Modular invariance reads here: $(S^J)^2=(-1)^p i \cdot 1=(S^JT^J)^3$ for $J=(1,0)$ and $(S^J)^2=(-1)^{p-1} i \cdot 1=(S^JT^J)^3$ for $J=(1,1)$, both with imaginary $(S^J)^2$.} combinations are in fact\rlap:\footnote{Note that in order to use these relations one must order the six fields as indicated above, without paying attention to the actual labelling of the fixed point fields.}
\begin{equation}
\label{S10p2bis}
\boxed{S^{(1,0)}= e^{-\frac{i\pi}{4}} \cdot (-i)^{p-1} \cdot S^{(1,1)}_{D4}
= e^{-\frac{i\pi}{4}(m-2)} \cdot S^{(1,1)}_{D4}}
\end{equation}
and
\begin{equation}
\label{S11p2bis}
\boxed{S^{(1,1)}= e^{-\frac{i\pi}{4}} \cdot (-i)^{p-1} \cdot S^{(1,0)}_{D4}
= e^{-\frac{i\pi}{4}(m-2)} \cdot S^{(1,0)}_{D4}}
\end{equation}
giving also acceptable fusion rules. Here $m=2p+1$ is an odd integer such that $D_{2m}\equiv D_{4p+2}$.

There are a few comments that we can make here. The first comment regards the labelling of the matrices just given. We observe that the matrix $S^{(1,0)}$ (resp. $S^{(1,1)}$) contains the same fields as the matrix $S^{(1,1)}_{D4}$ (resp. $S^{(1,0)}_{D4}$) except for the fact that the twisted fields corresponding to the spinors are interchanged (but they still have the same weights). We will then keep the same labels as given in the above scheme (\ref{D 4p+2 scheme}) and in table \ref{table S^J=11_p=1} (resp. table \ref{table S^J=10_p=1}).

The second comment regards the periodicity of the modular matrices. Observe that in (\ref{D 4p+2 scheme}) a shift $n\rightarrow n+16$ (corresponding to $m\rightarrow m+8$ and $p\rightarrow p+4$) changes all the weights by integers, but the $T^J$ matrices will be invariant. Similarly, the $S^J$ matrices are invariant under the same shift $m\rightarrow m+8$. This happened already for the modular matrices in the $n=4p$ case and it happens here again in the $n=4p+2$ case. Hence, it seems that in comparing phases one should consider situations which have the same $p$ mod $4$. On the other hand, in going from $n=4p$ to $n=4p+2$, the $S^J$ formulas are similar, but there is one main difference, namely $S^{(1,0)}_{D4}$ gets interchanged by $S^{(1,1)}_{D4}$ and this is a completely different matrix. The same consideration that we made after (\ref{S10p}) about the shift $p\rightarrow p+1$ can be repeated here.

The last comment regards the fusion coefficients. Note that when we check the fusion rules, we cannot do it directly from the single $D(n)_1$ permutation orbifolds, exactly because the spinor currents have half-integer spin. Instead, we have to tensor the $D(n)_1$ theory with another one which also has half-integer spin simple currents (e.g. Ising model or the $D(n)_1$ model itself, maybe with different values of $n$) such that the tensor product has integer spin simple currents that can be used for the extension: those integer spin currents will then have acceptable fusion coefficients. We have checked that this is indeed the case for tensor products of the permutation orbifold CFT's with the Ising model, and also in extensions of different permutation orbifold CFT's tensored with each other (we have also performed the latter check for $n=4p$, for combinations of integer spin currents).

\section{Conclusion}
In this paper we have completed the analysis initiated in \cite{Maio:2009kb} of extensions of $D(n)_1$ permutation orbifolds by additional integer spin simple currents arising when the rank $n$ is multiple of four and by additional half-integer spin simple currents arising when the rank $n$ is even but not multiple of four. In both situations fixed points occur that must be resolved in the extended theory. This means that we have to provide the $S^J$ matrices corresponding to those extra currents $J$. They will allow us to obtain the full $S$ matrix of the extended theory which satisfies all the necessary properties. 

The currents in question are those corresponding to the spinor representations $i=1$ and $i=3$ of $D(n)_1$, both with weight $h=\frac{n}{8}$. In the permutation orbifold they arise from the symmetric and the anti-symmetric representations of the spinors, both with weight $h=\frac{n}{4}$: so they have integer spin for $n=4p$ ($p$ is integer) and half-integer spin for $n=4p+2$. Moreover, they produce pairwise identical extensions of the permutation orbifold, such that there are only two unknown matrices to determine: $S^{(1,\psi)}=S^{(3,\psi)}$ ($\psi=0,1$). The solutions were given in sections \ref{D4p permutation orbifolds} and \ref{D4p+2 permutation orbifolds} (boxed formulas). This completely solves the fixed point resolution in extension of $D(n)_1$ permutation orbifold.

There is still more work to do. First of all, we do not have any general expression yet for the $S^J$ matrix in terms of the $S$ (and maybe $P$) matrix of the mother theory. This should be independent of the particular CFT and/or the particular current used to extend the theory. Secondly, it would be interesting to apply these CFT results in String Theory. Suitable candidates appear to be the minimal models of the $N=2$ superconformal algebra, which are the building blocks of Gepner models \cite{Gepner:1987vz,Gepner:1987qi}, but this is still work in progress.

\section*{Acknowledgments}
This research is supported by the Dutch Foundation for Fundamental Research of Matter (FOM)
as part of the program STQG (String Theory and Quantum Gravity, FP 57) and has been partially supported by funding of the Spanish Ministerio de Ciencia e Innovaci\'on, Research Project FPA2008-02968.

\end{document}